\documentclass[onecollarge,natbib,sort&compress]{svjour2}
\bibpunct{[}{]}{;}{n}{}{,} % to get "[numbered]" references from natbib
\smartqed  % flush right qed marks, e.g. at end of proof
\usepackage{graphicx}
\usepackage[normalem]{ulem}
 \usepackage{amsmath}
\usepackage{amssymb}
\usepackage{mathtools}
\usepackage{color}
\usepackage{xcolor}
\usepackage[pdftex,hypertexnames=false,linktocpage=true]{hyperref}
\hypersetup{colorlinks=true,linkcolor=blue,anchorcolor=blue,citecolor=darkgreen, filecolor=blue,urlcolor=blue,bookmarksnumbered=true,
pdfview=FitB
}
\usepackage{lineno}
\usepackage{relsize}
\usepackage{mathrsfs}
\usepackage{bm}
\usepackage{subfigure}
\usepackage{cancel}
\usepackage{soul}
\usepackage{booktabs}
\usepackage{multirow}

\colorlet{darkgreen}{green!50!black}
\colorlet{brightyellow}{yellow!75!red}
\colorlet{orange}{red!50!yellow}
\colorlet{darkblue}{blue!60!black}
\colorlet{darkred}{red!80!black}

\newcommand{\dd} {{\mathrm{d}}}

\newcommand{\half}[1][1] {{\mathsmaller{\frac{#1}{2}}}}
\journalname{}
\begin{document}
 
\title{Quarkonium wave functions on the light front}
% \subtitle{Do you have a subtitle?\\ If so, write it here}

\titlerunning{quarkonium}        % if too long for running head

\author{Yang~Li}

\authorrunning{Yang Li} % if too long for running head

\institute{Yang Li \at
              Department of Physics and Astronomy, Iowa State University, Ames, IA 50011, USA \\
              \email{leeyoung@iastate.edu}
}

\date{\today}

\maketitle

\begin{abstract}
We study heavy quarkonium within the light-front Hamiltonian formalism. Our effective Hamiltonian is based on 
the holographic QCD confining potential and the one-gluon exchange interaction with a running coupling. 
The obtained spectra are compared with experimental measurements. We present a set of light-front 
wave functions, which exhibit rich structure and are consistent with the nonrelativistic picture. Finally, we use the
wave functions to compute the charge and mass radii. 
\end{abstract}

%\linenumbers

%%%%%%%%%%%%%%%%%%%%%%
\section{Introduction}\label{introduction}

 The light-front Hamiltonian approach is poised as a unique tool to understand the non-perturbative dynamics
 of quantum field theory in Minkowski space-time \cite{Bakker13}. The Hamiltonian formalism 
 provides frame-independent wave functions that are essential for understanding the underlying structure of  
 relativistic bound state systems. The obtained light-front wave functions (LFWFs) can be used to compute 
 hadronic observables and distributions defined in the infinite momentum frame (IMF), and to study exclusive 
 processes in the deep inelastic scattering (DIS), which are otherwise not easily accessible in other methods \cite{Lepage80}. 
 Therefore, the light-front Hamiltonian approach is complementary to Lagrangian methods formulated in 
 Euclidean space, e.g., lattice gauge theory. 

 Recently we proposed a phenomenological light-front potential model for mesons based on light-front holography
  \cite{Li16}. We applied this model to charmonium and bottomonium, and solved the system in a basis function 
  representation. The obtained mass spectra, form factors and decay constants are compared with experiments 
  and other established methods. The LFWFs are also used to study exclusive vector meson productions in DIS \cite{Chen16}.
  Very recently, similar work has been done within the Dyson-Schwinger/Bethe-Salpeter approach \cite{Hilger15} and 
  the covariant spectator theory (CST) \cite{Leitao17}. 
 
 In this paper, we propose an improvement of the model in Ref.~\cite{Li16} by incorporating the evolution of the coupling based on perturbative 
 QCD (pQCD). We will show that the mass spectra are improved. Another aim of this paper is to display the LFWFs,
 which reveal first-hand information of the hadrons, and also help to visualize relativistic bound states on the light front. 
Heavy quarkonium is an ideal system to explore LFWFs, as the characteristic heavy quark velocities are small [$v\sim(0.1\text{--}0.3)$] compared 
to the speed of light  and the familiar nonrelativistic 
quantum mechanical language and intuition may be applied. 

%%%%%%%%%%%%%%%%%%%%%%%%%%%%%%%%
\section{Formalism} \label{sec 2}

The effective Hamiltonian for the model reads \cite{Li16}, 
\begin{equation}\label{eqn:Heff}
\begin{split}
H_\mathrm{eff} &= \frac{\vec k^2_\perp + m_q^2}{x} + \frac{\vec k^2_\perp+m_{\bar q}^2}{1-x}
+ \kappa^4 \vec \zeta_\perp^2 - \frac{\kappa^4}{(m_q+m_{\bar q})^2} \partial_x\big( x(1-x) \partial_x \big)  \\
&\quad - \frac{C_F4\pi\alpha_s(Q^2)}{Q^2}\bar u_{s'}(k')\gamma_\mu u_s(k) \bar v_{\bar s}(\bar k) \gamma^\mu v_{\bar s'}(\bar k').
\end{split}
\end{equation}
where $Q^2 = -\bar q^2 = -(1/2) (k'-k)^2- (1/2) (\bar k'-\bar k)^2$ is the average 4-momentum squared carried by 
the exchanged gluon. 
The confining term $\kappa^4 \vec\zeta^2_\perp$ comes from the ``soft-wall'' light-front holography, $\kappa$ 
is the strength of the confinement, and $\vec \zeta_\perp \equiv \sqrt{x(1-x)} \vec r_\perp$ is Brodsky and de T\'eramond's 
holographic variable \cite{Brodsky15}. $\vec \zeta_\perp$ and $x$ are taken as the independent variables, 
 $\partial_x f(x, \vec\zeta_\perp) = \partial f(x, \vec \zeta_\perp)/\partial x|_{\vec\zeta_\perp}$. $C_F = (N_c^2-1)/(2N_c)=4/3$ 
 is the color factor. $m_q$ ($m_{\bar q}$) is the mass of the quark (anti-quark). 
 The longitudinal confinement comes from completing the transverse holographic confinement in the finite mass case and is 
 consistent with the pQCD asymptotics of the distribution amplitude $x^\beta(1-x)^\alpha$. The longitudinal confining strength 
 is fixed in the nonrelativistic limit hence no additional parameter is introduced. 

In Ref.~\cite{Li16}, the strong coupling $\alpha_s$ is assumed fixed for each system, say, charmonium, whereas for different 
systems, the values of $\alpha_s$ are different and are related through the pQCD evolution. While such treatment is reasonable 
(indeed traditional, see e.g. \cite{Griffiths08}), as the mass difference between states in the same system, e.g. $J/\psi$ and 
$\psi(2S)$, is small compared to their respective masses, nevertheless, the inclusion of the running coupling implements 
important QCD physics. 
Another motivation is to improve the rotational symmetry and the hyperfine structure. In the effective one-gluon exchange kernel, 
\cite{Krautgartner92,Lamm14, Wiecki15}, a non-covariant UV counterterm is introduced to regularize the UV divergence appearing
in the box diagram \cite{Krautgartner92, Mangin-Brinet03}, which further spoils the rotational symmetry as will be seen in the mass
 spectra below.
The introduction of the running coupling changes the UV asymptotics of the one-gluon exchange kernel~\footnote{This is not to say, 
however, that now the UV asymptotics of the one-gluon exchange kernel is consistent with 
the perturbation theory. But the UV asymptotics is not easily studied in the basis representation and the investigation of the UV asymptotics 
is beyond the scope of this paper.
}. As a result, the counterterm is not necessary anymore. While rotational symmetry is not fully restored, as we will see, the 
breaking of rotational symmetry is moderated, and the hyperfine structure is significantly improved. 

In the effective one-gluon exchange interaction Eq.~(\ref{eqn:Heff}), the 4-momentum squared of the exchanged gluon $Q^2$ 
naturally appears. The running coupling can be introduced as a function of $Q^2$: 
\begin{equation}\label{eqn:running_couplings}
\alpha_s(Q^2) =
\frac{\alpha_s(M_\textsc{z}^2)}{1+\alpha_s(M_\textsc{z}^2)\beta_0
\ln(\mu^2_\textsc{ir}+Q^2)/(\mu^2_\textsc{ir}+M^2_\textsc{z})}.
\end{equation} 
where $N_f = 4$ for charmonium and $N_f=5$ for bottomonium. 
Here $\mu_\textsc{ir}$ is an IR cutoff introduced to avoid the pQCD IR catastrophe.
Similar forms are used in the literature, e.g. Ref.~\cite{Atkinson88}.  
In practice, we choose $\alpha_s(0)=0.6$. We find the spectrum is not
sensitive for the choice of $\alpha_s(0)$ within the range $0.4 \le \alpha_s(0) \le 0.8$.

The state vectors, hence the LFWFs, are obtained from diagonalizing the Light Cone Hamiltonian operator $\hat H_\textsc{lc}
\equiv P^+ \hat P^- - \vec P^2_\perp$: 
\begin{equation}\label{eqn:eigenvalue_equation}
H_\textsc{lc} |\psi_h(P^+, \vec P_\perp, j, m_j)\rangle = M^2_h |\psi_h(P^+, \vec P_\perp, j, m_j)\rangle.
\end{equation}
The state vectors can be represented in the Fock space:
\begin{multline}\label{eqn:Fock_expansion}
 |\psi_h(P^+, \vec P_\perp, j, m_j)\rangle = 
\sum_{s, \bar s}\int_0^1\frac{\dd x}{2x(1-x)} \int \frac{\dd^2 \vec k_\perp}{(2\pi)^3}
\, \psi^{(m_j)}_{s\bar s/h}(\vec k_\perp, x) \\
\times \frac{1}{\sqrt{N_c}}\sum_{i=1}^{N_c} b^\dagger_{s{}i}(xP^+, \vec k_\perp+x\vec P_\perp) d^\dagger_{\bar s{}i}((1-x)P^+, -\vec 
k_\perp+(1-x)\vec P_\perp) |0\rangle. 
\end{multline}
Here $j$ and $m_j$ are the intrinsic spin and the spin projection of the particle. The coefficient of the expansion, $\psi^{(m_j)}_{s\bar
s/h}(\vec k_\perp, x)$, is the (two-body) LFWF, which is normalized as, 
\begin{equation}\label{eqn:normalization}
 \sum_{s, \bar s} \int_0^1\frac{\dd x}{2x(1-x)} \int \frac{\dd^2 \vec k_\perp}{(2\pi)^3} 
\psi^{(m_j')*}_{s \bar s/h'}(\vec k_\perp, x)\psi^{(m_j)}_{s \bar s/h}(\vec k_\perp, x)  = \delta_{hh'}\delta_{m_j,m_j'}.
\end{equation}
It is useful to introduce LFWFs in the transverse coordinate space:
\begin{equation}
  \widetilde\psi_{s\bar s} (\vec r_\perp, x) \equiv \frac{1}{\sqrt{x(1-x)}}\int \frac{\dd^2 k_\perp}{(2\pi)^2}  e^{i \vec k_\perp \cdot
\vec
r_\perp}
\psi_{s \bar s} (\vec k_\perp, x).
\end{equation} 

Following Ref.~\cite{Li16}, the effective Hamiltonian Eq.~(\ref{eqn:Heff}) was solved in a basis function approach. The LFWFs then are represented as, 
\begin{equation}\label{eqn:basis_representation}
 \begin{split}
 \psi_{ss'/h}(\vec k_\perp, x) =& \sum_{n, m, l} \psi_h(n, m, l, s, s') \, \phi_{nm}(\vec k_\perp/\sqrt{x(1-x)}) \chi_l(x); \\
 \widetilde\psi_{ss'/h}(\vec r_\perp, x) =& \sqrt{x(1-x)}\sum_{n, m, l} \psi_h(n, m, l, s, s') \, \widetilde\phi_{nm}(\sqrt{x(1-x)}\vec
r_\perp) \chi_l(x).\\
 \end{split}
\end{equation}
Here the coefficients $ \psi_h(n, m, l, s, s')$ are obtained from diagonalization. Basis function  
$\phi_{nm}$ and $\chi_l$ are the solutions of the Hamiltonian without the one-gluon exchange 
interaction, i.e., the kinetic energy plus the confinement terms. Specifically, $\widetilde\phi_{nm}(\vec \zeta_\perp)$ is the harmonic
oscillator function in the holographic variables. $\chi_\ell(x)$ is the Jacobi polynomials weighed by 
$x^{\half[\beta]}(1-x)^{\half[\alpha]}$, where $\alpha = 2m_{\bar q}(m_q+m_{\bar q})/\kappa^2$, $\beta = 2m_{q}(m_q+m_{\bar q})/\kappa^2$.

In practical calculations, the basis is truncated and LFWFs have a corresponding finite set of terms in Eqs.~(\ref{eqn:basis_representation}). Observables are evaluated
from finite-dimensional representations and, in principle, have to be extrapolated to the complete basis. Following Ref.~\cite{Li16}, the basis are
truncated according to:
\begin{equation}
 2n + |m| + 1 \le N_\mathrm{max}, \quad 0 \le l \le L_{\mathrm{max}}.
\end{equation}

\section{Results}

\begin{table}
\caption{Summary of the model parameters.  
$\mu_g$ is the gluon mass, used to regularize the integrable Coulomb singularity. 
$\langle M_\text{cal}-M_\text{pdg}\rangle$ represents the r.m.s. deviation of the masses from the Particle Data Group 
values for states below the threshold. 
$\langle\delta M_{m_j}\rangle$ represents the r.m.s. value of the mass spreads for the same state with different magnetic
projection $m_j$'s. 
}\label{tab:model_parameters}
 \centering
\begin{tabular}{ccc ccc ccc} 

\toprule 

   & $N_f$ & $\mu_\text{g}$ (GeV) & $\kappa$ (GeV) & $m_q$ (GeV) & $\langle M_\text{cal}-M_\text{pdg}\rangle$ & $\langle\delta M_{m_j}\rangle$ & \#~states &
$N_{\max}=L_{\max}$ \\ %$\delta M_{m_J=0}$

\midrule

 \multirow{4}{*}{$c\bar c$}  & \multirow{4}{*}{4} & \multirow{4}{*}{0.02} & 0.985 & 1.570 & 41 MeV  & 15 MeV &  \multirow{4}{*}{8 states} & 8 \\ %max: 23.4 MeV (8 states), mean: 11.5 MeV (8 states), 14.5 MeV (5 states, j>0) %42 MeV
     &                              &   & 0.979 & 1.587 & 32{\qquad\;\;}  & 21{\qquad\;\;}  & &16 \\ %max: 31.9 MeV (8 states), mean: 16.7 MeV (8 states), 21.1 MeV (5 states, j>0) %35  MeV
    &                              &  & 0.972 & 1.596 & 31{\qquad\;\;} & 17{\qquad\;\;}  & & 24 \\ %max: 29.2 MeV (8 states), mean: 13.2 MeV (8 states), 16.6 MeV (5 states, j>0) %33 MeV
   &                              &  & 0.966 & 1.603 & 31{\qquad\;\;}   & 17{\qquad\;\;}  & & 32 \\ %max: 30.467 MeV (8 states), mean: 13.5 MeV (8 states), 17.1 MeV (5 states, j>0) %33 MeV 
\midrule 
  \multirow{4}{*}{$b\bar b$} & \multirow{4}{*}{5} & \multirow{4}{*}{0.02} & 1.387 & 4.894 & 48 MeV & 6 MeV & \multirow{4}{*}{14 states} & 8 \\ %max: 12.3 MeV (30 states), mean: 2.1 MeV (14 states), 5.7 MeV (25 states, j>0) %48 MeV 
   &                                  & & 1.392 & 4.899 & 41{\qquad\;\;}  &  6{\qquad\;\;}  & & 16 \\  %max: 12.6 MeV (30 states), mean: 2.8 MeV (14 states), 6.1 MeV (25 states, j>0) %41 MeV
  &                                   & & 1.390 & 4.901 & 39{\qquad\;\;}  & 6{\qquad\;\;}  & & 24 \\  %max: 12.5 MeV (30 states), mean: 3.1 MeV (14 states), 6.2 MeV (25 states, j>0) %39 MeV
  &                                   & & 1.389 & 4.902 & 38{\qquad\;\;}  & 8{\qquad\;\;}  & & 32 \\ %max: 12.4 MeV (30 states), mean: 3.3 MeV (14 states), 6.2 MeV (25 states, j>0) %38 MeV
\bottomrule  
\end{tabular}
\end{table}

\begin{figure}
\centering 
\includegraphics[width=0.45\textwidth]{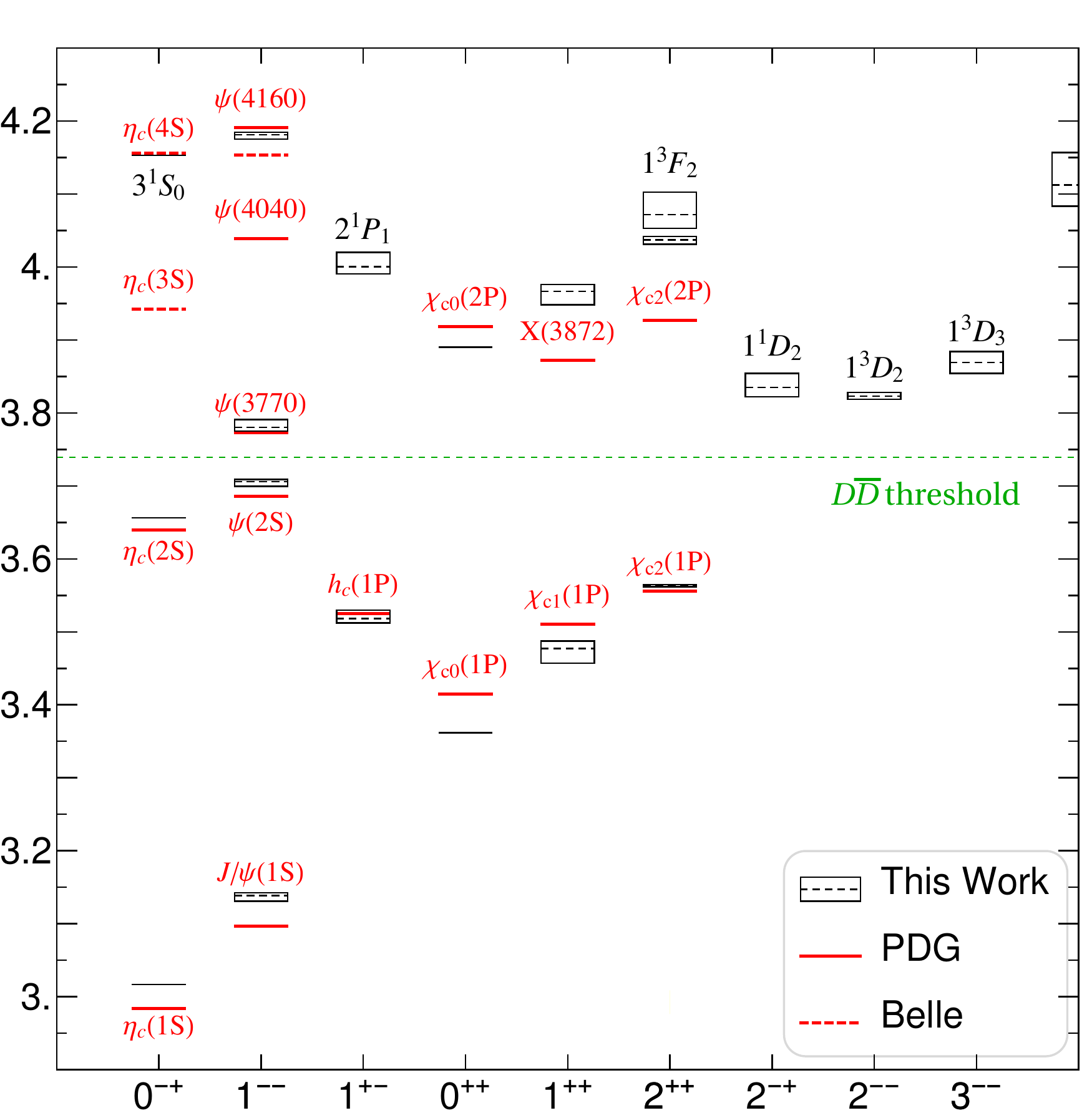} \quad
\includegraphics[width=0.46\textwidth]{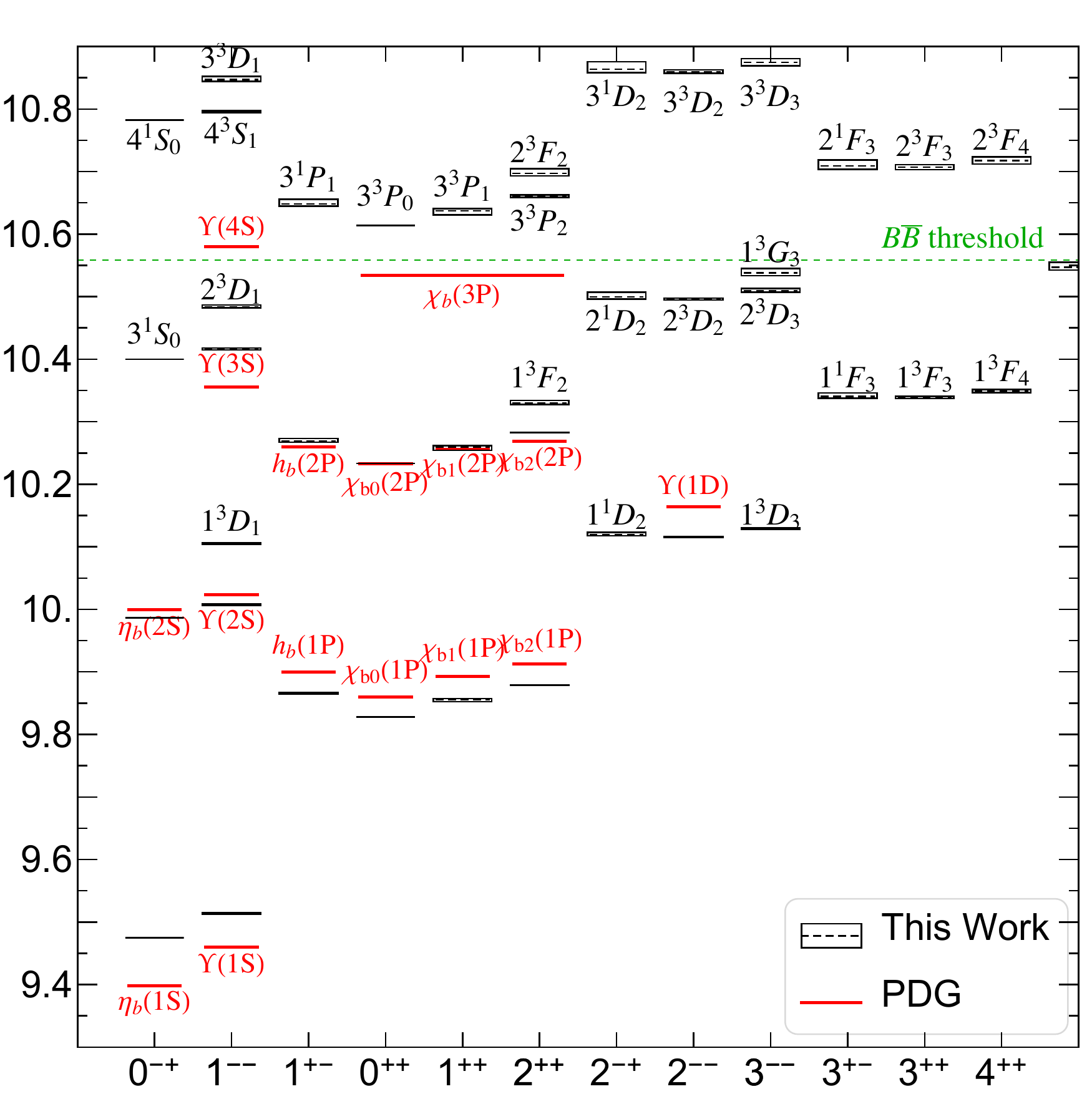}
\caption{The reconstructed charmonium (\textit{left}) and bottomonium (\textit{right}) spectra at $N_{\max}=L_{\max}=32$. 
The horizontal axes marks the quantum numbers $J^\mathrm{PC}$. The vertical axes marks the mass eigenvalues in GeV.
The parameters are listed in Table~\ref{tab:model_parameters}. The r.m.s. deviation of the masses from the 
PDG values are 31 MeV and 38 MeV for charmonium and bottomonium, respectively. The mean mass spreads, 
i.e. the mean heights of the boxes, are 17 MeV and 8 MeV for charmonium and bottomonium, respectively.}
\label{fig:spectroscopy}
\end{figure}

The model parameters are summarized in Table~\ref{tab:model_parameters}. In particular, the confining 
strength $\kappa$ and the effective quark mass $m_q$ are obtained from fitting to the PDG mass spectrum
below the continuum thresholds. 
The reconstructed charmonium and bottomonium spectra are shown in Fig.~\ref{fig:spectroscopy}.
 The r.m.s. deviation of the masses from the PDG values are 31 MeV and 38 MeV for charmonia and bottomonia below the
 thresholds, respectively. These r.m.s. deviations are significantly reduced 
 ($\sim$50\% for charmonium, $\sim$25\% for bottomonium) from the fixed coupling results reported in Ref.~\cite{Li16}. 
 Our spectroscopy is competitive with those obtained from other methods, e.g. \cite{Crater10, Hilger15, Leitao17}.
Not only are the mass eigevalues improved, their dependence on $m_j$, an indicator for the violation of the 
rotational symmetry, is also weakened. 
For example, the mean mass spreads, i.e. the mean heights of the boxes in Fig.~\ref{fig:spectroscopy}, are 17 MeV and 8 MeV for 
charmonia and bottomonia below the thresholds, respectively.

\begin{figure}
%\centering 
\subfigure[\ $\widetilde\psi_{\uparrow\downarrow+\downarrow\uparrow}^{\lambda=0}(\vec r_\perp, x)$ \label{fig:jpsi.mj0.triplet}]{
\includegraphics[width=0.32\textwidth]{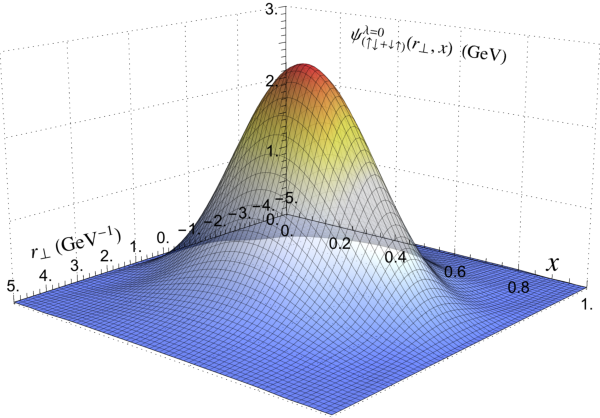}
}
\subfigure[\ $\widetilde\psi_{\downarrow\downarrow}^{\lambda=0}(\vec r_\perp, x)$  \label{fig:jpsi.mj0.mm}]{
\includegraphics[width=0.32\textwidth]{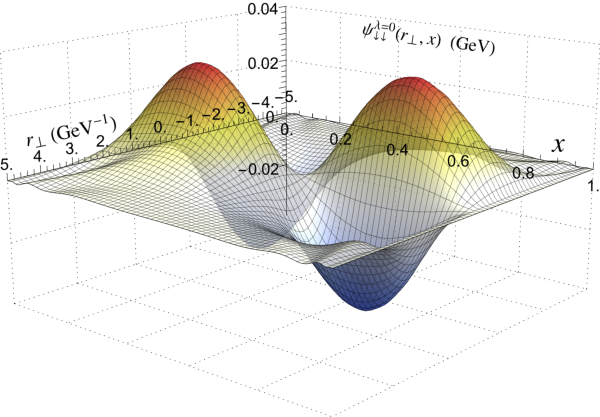}
}
\subfigure[\ $\widetilde\psi_{\uparrow\downarrow-\downarrow\uparrow}^{\lambda=+1}(\vec r_\perp, x)$ \label{fig:jpsi.mj1.singlet}]{
\includegraphics[width=0.32\textwidth]{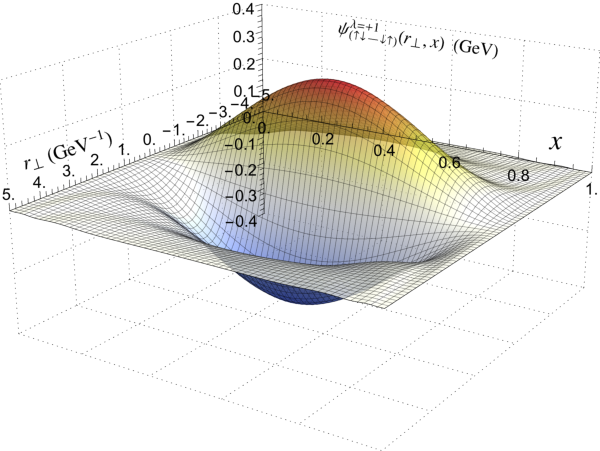}
}

\subfigure[\ $\widetilde\psi_{\uparrow\uparrow}^{\lambda=+1}(\vec r_\perp, x)$  \label{fig:jpsi.mj1.pp}]{
\includegraphics[width=0.32\textwidth]{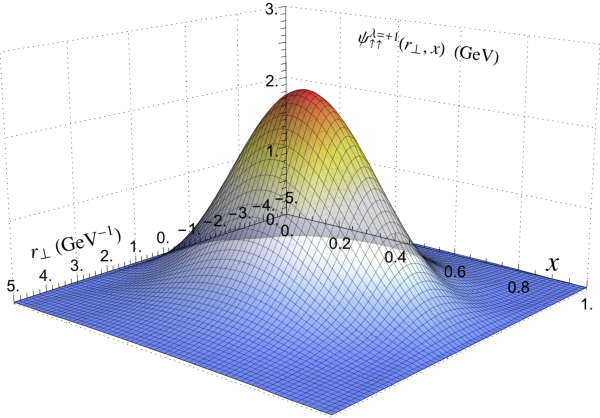}
}
\subfigure[\ $\widetilde\psi_{\uparrow\downarrow+\downarrow\uparrow}^{\lambda=+1}(\vec r_\perp, x)$ \label{fig:jpsi.mj1.triplet}]{
\includegraphics[width=0.32\textwidth]{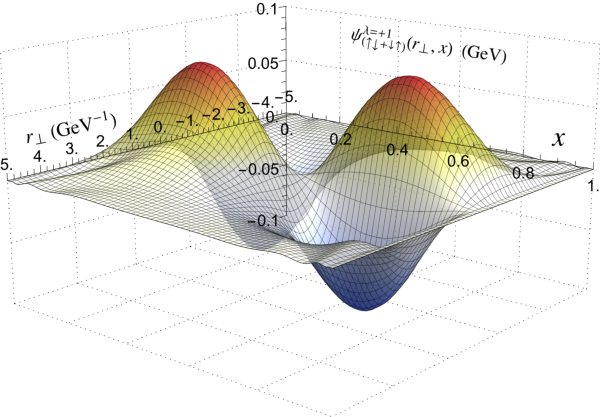}
}
\subfigure[\ $\widetilde\psi_{\downarrow\downarrow}^{\lambda=+1}(\vec r_\perp, x)$ \label{fig:jpsi.mj1.mm}]{
\includegraphics[width=0.32\textwidth]{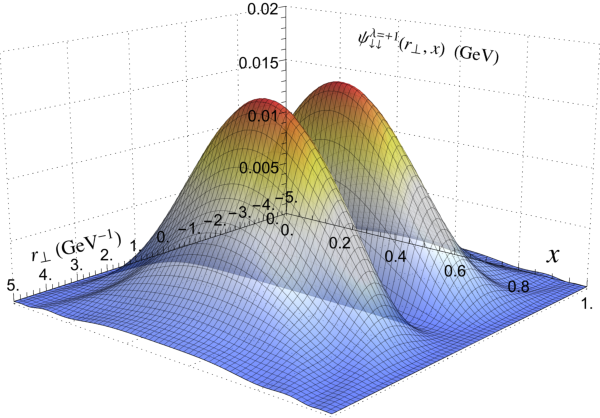}
}
\caption{(Color online) The $J/\psi$ LFWFs plotted in 3D for polarizations $\lambda=0$  (\emph{a--b}), and $\lambda=+1$ (\emph{c--f}), for various spin alignments.}
\label{fig:jpsi}
\end{figure}

\begin{figure}
%\centering 
\subfigure[\ $J/\psi$: $\widetilde\psi_{\uparrow\downarrow+\downarrow\uparrow}^{\lambda=0}(\vec r_\perp, x)$ \label{fig:jpsi.mj0.triplet.den}]{
\includegraphics[width=0.32\textwidth]{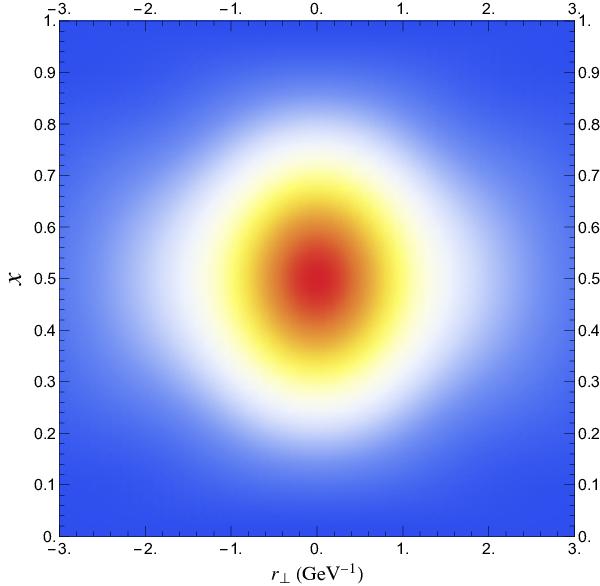}
}
\subfigure[\ $\psi(2S)$: $\widetilde\psi_{\uparrow\downarrow+\downarrow\uparrow}^{\lambda=0}(\vec r_\perp, x)$  \label{fig:psi2S.mj0.triplet.den}]{
\includegraphics[width=0.32\textwidth]{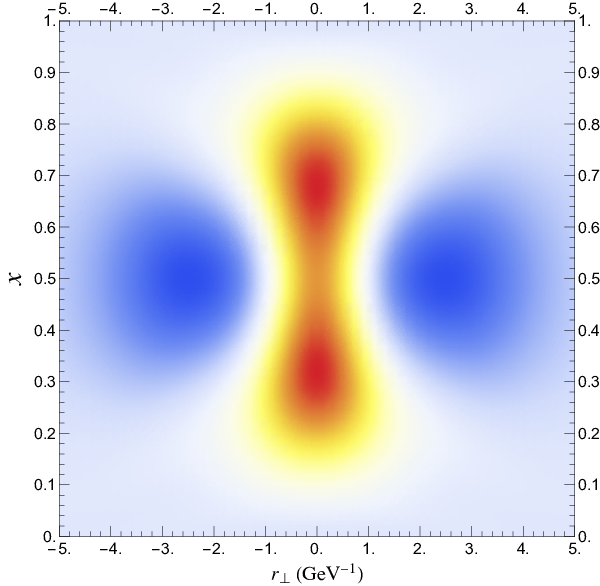}
}
\subfigure[\ $\psi(1D)$: $\widetilde\psi_{\uparrow\downarrow+\downarrow\uparrow}^{\lambda=0}(\vec r_\perp, x)$  \label{fig:psi1D.mj0.triplet.den}]{
\includegraphics[width=0.32\textwidth]{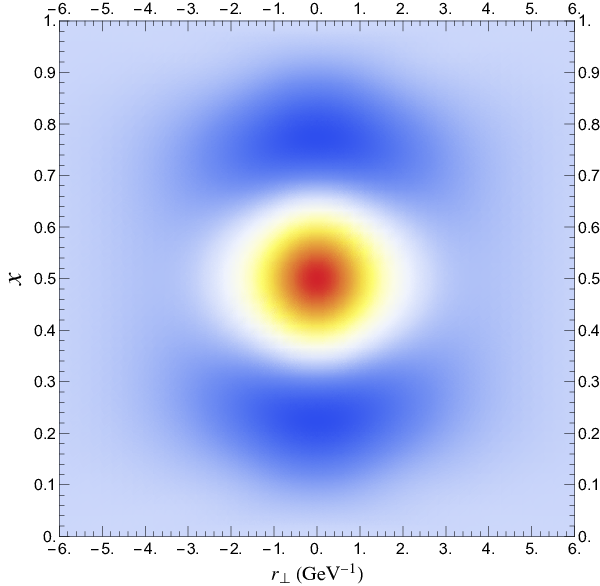}
}

\subfigure[\ $J/\psi$: $\widetilde\psi_{\uparrow\uparrow}^{\lambda=+1}(\vec r_\perp, x)$  \label{fig:jpsi.mj1.pp.den}]{
\includegraphics[width=0.32\textwidth]{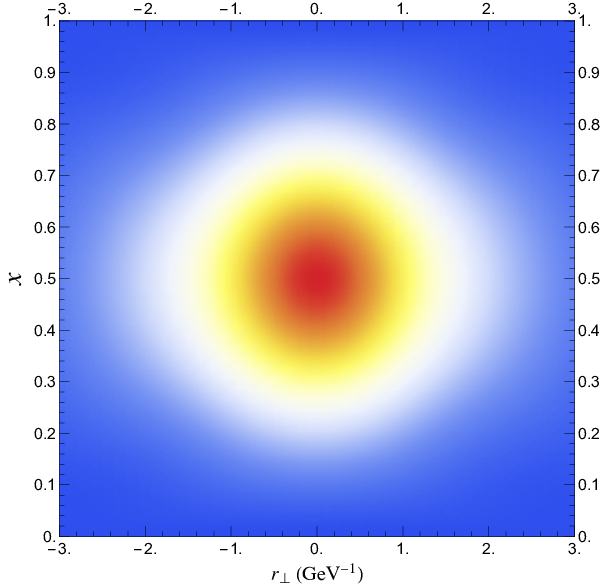}
}
\subfigure[\ $\psi(2S)$: $\widetilde\psi_{\uparrow\uparrow}^{\lambda=+1}(\vec r_\perp, x)$  \label{fig:psi2S.mj1.pp.den}]{
\includegraphics[width=0.32\textwidth]{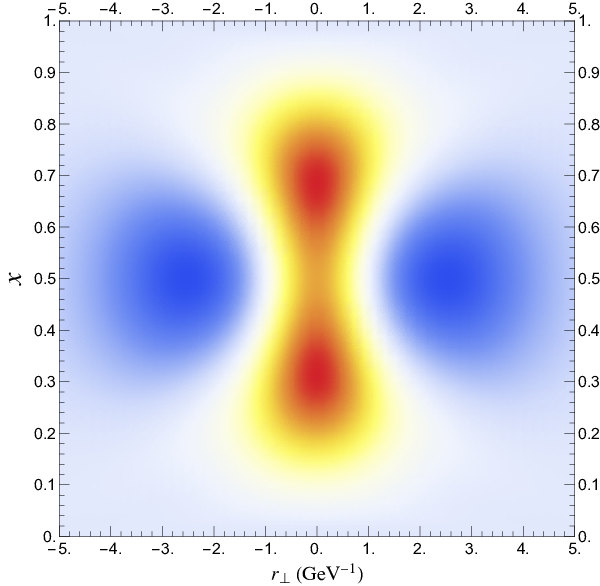}
}
\subfigure[\ $\psi(1D)$: $\widetilde\psi_{\downarrow\downarrow}^{\lambda=+1}(\vec r_\perp, x)$ \label{fig:psi1D.mj1.mm.den}]{
\includegraphics[width=0.32\textwidth]{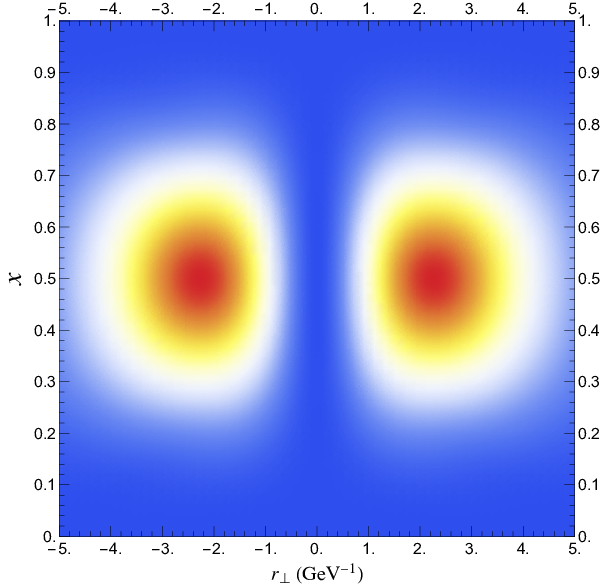}
}
\caption{(Color online) Leading components of the vector meson $J/\psi$ and its radial excitation $\psi(2S)$ and angular excitation $\psi(1D)$
for polarization $\lambda=0$ (\emph{a--c}) and $\lambda=+1$ (\emph{d--f}). The color scheme is the same as Fig.~\ref{fig:jpsi}.
}
\label{fig:vm}
\end{figure}

\begin{figure}
\centering 
\subfigure[\ $\widetilde\psi_{\uparrow\downarrow-\downarrow\uparrow}^{\lambda=0}(\vec r_\perp, x)$ \label{fig:etab3S.mj0.triplet.den}]{
\includegraphics[width=0.32\textwidth]{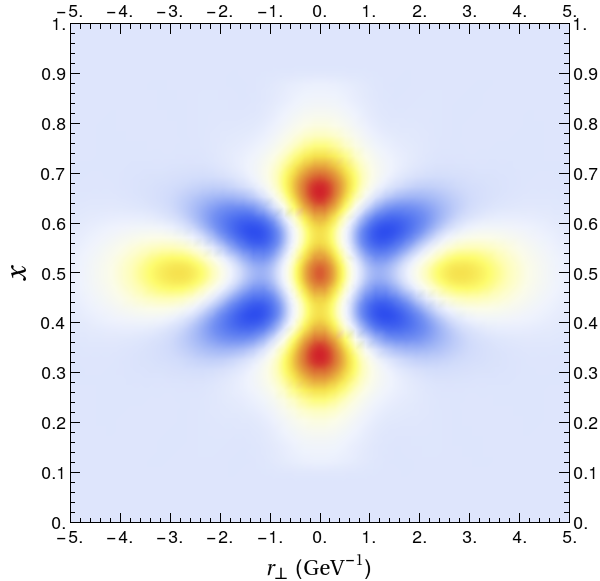}
}
\quad
\subfigure[\ $\widetilde\psi_{\downarrow\downarrow}^{\lambda=0}(\vec r_\perp, x)$ \label{fig:etab3S.mj0.mm.den}]{
\includegraphics[width=0.32\textwidth]{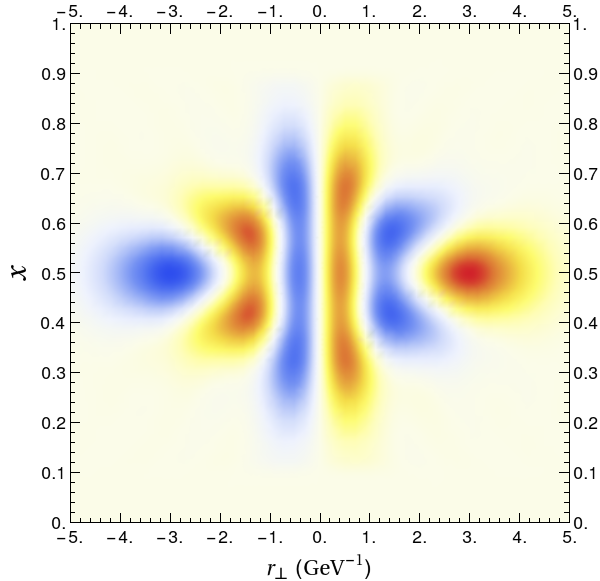}
}
\caption{(Color online) LFWFs of $\eta_b(3S)$. The color scheme is the same as Fig.~\ref{fig:jpsi}.}
\label{fig:etab3S}
\end{figure}

Wave functions provide full information of the system. Despite of their significance in hadron physics, 
LFWFs, especially those of the excited states, are rarely found in the literature (cf.~\cite{Glazek93}). 
We present representative charmonium LFWFs in Figs.~\ref{fig:jpsi}--\ref{fig:etab3S}.

Figure~\ref{fig:jpsi} shows the coordinate space LFWFs for vector meson $J/\psi$. For each spin alignment, 
the orbital angular momentum projection $m_\ell=\lambda-s_1-s_2$ is definite ($\lambda\equiv m_j$). Hereafter, we drop the global phases
$\exp(i m_\ell \theta)$ related to the orbital angular momentum, while retaining the relative sign for negative values
of $r_\perp$. 
It is apparent that $\psi^{\lambda=0}_{\uparrow\downarrow+\downarrow\uparrow}$ [Fig.~\ref{fig:jpsi.mj0.triplet}] and
$\psi^{\lambda=+1}_{\uparrow\uparrow}$ [Fig.~\ref{fig:jpsi.mj1.pp}] are the leading LFWFs for polarization 
$\lambda=0$ and $\lambda=+1$, respectively. And they are almost identical in magnitude. Thus the nonrelativistic picture 
emerges that $J/\psi$ is an $S$-wave in direct product with a triplet spin configuration. 
Subleading LFWFs also exist [Figs.~\ref{fig:jpsi.mj0.mm}, \ref{fig:jpsi.mj1.singlet}, \ref{fig:jpsi.mj1.triplet}, \ref{fig:jpsi.mj1.mm}] 
due to relativity. For example, $\psi^{\lambda=+1}_{\downarrow\downarrow}$ 
[Fig.~\ref{fig:jpsi.mj1.mm}] resembles a $D$-wave function ($\ell=2, m_\ell= 2$), as a result of the $S$-$D$ mixing.

Figure~\ref{fig:vm} compares the leading components of the vector meson $J/\psi$ with its ``radial'' excitation $\psi(2S)$ and
``angular'' excitation $\psi(1D)$. 
Rich details emerge in the excited states.
The $\psi(2S)$ LFWFs show nodes in both the transverse radial direction ($r_\perp$) and the longitudinal direction ($x$), consistent with the
nonrelativistic interpretation, where the radial excitation is homogeneous in all 3 directions.   
LFWFs on the top panel (bottom panel) are with the polarization $\lambda=0$ ($\lambda=+1$).
While the leading components of a $S$-wave vector meson, e.g., $J/\psi$ and $\psi(2S)$, for the all polarizations are almost identical
[Fig.~\ref{fig:jpsi.mj0.triplet.den}~vs~\ref{fig:jpsi.mj1.pp.den} and  Fig.~\ref{fig:psi2S.mj0.triplet.den}~vs~\ref{fig:psi2S.mj1.pp.den}], 
those of the $D$-wave vector meson $\psi(1D)$ are dramatically different [Fig.~\ref{fig:psi1D.mj0.triplet.den} vs \ref{fig:psi1D.mj1.mm.den}]. 
However, this is consistent with the nonrelativistic picture  that the wave function is dominated by the $D$-wave ($\ell=2$).  
It is easy to see that the leading components of $\psi(1D)$ are $\widetilde\psi_{\uparrow\downarrow+\downarrow\uparrow}^{\lambda=0}(\vec r_\perp, x)$ ($m_\ell = 0$) 
and $\widetilde\psi_{\downarrow\downarrow}^{\lambda=+1}(\vec r_\perp, x)$ ($m_\ell=2$). 
The apparent difference between the two polarizations in fact resembles the difference between the $D$-wave spherical harmonics
$Y_2^0(\hat r)$ and $Y_2^2(\hat r)$.

Figure~\ref{fig:etab3S} presents the LFWFs of $\eta_b(3S)$. The wave functions for this highly excited state reveal 
complicated inner structure. The leading components $\psi_{\uparrow\downarrow-\downarrow\uparrow}^{\lambda=0}(\vec r_\perp, x)$
has additional nodes in both the transverse and the longitudinal directions compared with $2S$-wave states and is presumably
dominated by a $3S$-wave. The subleading LFWF $\psi_{\downarrow\downarrow}^{\lambda=0}(\vec r_\perp, x)$ also exhibits multiple 
nodes and may contain both radial and angular excitations.

Another way of visualizing relativistic bound states, in connection with the experiments, is to use the impact parameter generalized parton distributions (GPDs) \cite{Burkardt00}. These quantities are closely related to the LFWFs \cite{Diehl03}. In fact, they are just the squared LFWFs in the coordinate space, inserting gauge links as necessary. The impact parameter GPDs for a similar system---positronium---have been shown in Ref.~\cite{Adhikari16}.

\section{Charge and Mass Radii}

We use the LFWFs to study the the root-mean squared (r.m.s.) charge and the mass radii. In relativistic dynamics, the 
r.m.s. radii are defined from the corresponding form factors (electromagnetic form factor 
$F_c$ and the gravitational form factor $F_g$) which in turn are related
to the hadron matrix elements of conserved currents (electromagnetic current 
$J^\mu$ and the energy-momentum tensor $T^{\mu\nu}$):
\begin{equation}
 \langle r^2_c\rangle \equiv -6 \frac{\partial}{\partial  q^2} F_c(q^2)\big|_{q\to 0}, \qquad
  \langle r^2_m\rangle \equiv  -6 \frac{\partial}{\partial  q^2} F_g(q^2)\big|_{q\to 0}.
\end{equation}
For the charge form factor, we only couple the probing photon to the quark.
In light-front dynamics, it can be shown (see Appedix~\ref{app}) that the r.m.s. radii 
are related to the impact parameter $\vec b_\perp \equiv (1-x)\vec r_\perp$ \cite{Burkardt00} and the 
holographic variable $\vec \zeta_\perp \equiv \sqrt{x(1-x)} \vec r_\perp$ \cite{Brodsky15}:
\begin{align} 
&\langle r^2_c\rangle 
= \frac{3}{2} \sum_{s,s'} \int_0^1 \frac{\dd x}{4\pi} \int \dd^2r_\perp \, (1-x)^2\vec r^2_\perp \,
\widetilde\psi_{ss'/h}^{(\lambda)*}(\vec r_\perp, x)
\widetilde\psi_{ss'/h}^{(\lambda)}(\vec r_\perp, x) \equiv \frac{3}{2} \langle \vec b^2_\perp \rangle; \label{eqn:light-front_charge_radii}\\
&\langle r^2_m\rangle = \frac{3}{2} \sum_{s,s'} \int_0^1 \frac{\dd x}{4\pi} \int \dd^2r_\perp \, x(1-x)\vec r^2_\perp \,
\widetilde\psi_{ss'/h}^{(\lambda)*}(\vec r_\perp, x)
\widetilde\psi_{ss'/h}^{(\lambda)}(\vec r_\perp, x) \equiv \frac{3}{2} \langle \vec \zeta^2_\perp \rangle. \label{eqn:light-front_gravitational_radii}
 \end{align}
The results are summarized in Table~\ref{tab:radii}. Note that the difference between $\langle r_c^2 \rangle$ and $\langle r_m^2 \rangle$
is a relativistic effect.

\begin{table}
\centering
\caption{The r.m.s. charge and the mass radii of charmonia and bottomonia. Results are extrapolated 
from $N_{\max}=8,16,24,32$. Numerical uncertainties are estimated from the difference between 
the extrapolated values and the $N_{\max}=32$ values.
}
\label{tab:radii}
\begin{tabular}{ccccc ccccc}
\toprule
(fm) & $\eta_c$ & $\chi_{c0}$ & $\eta_c(2S)$ &  $\eta_b$ & $\chi_{b0}$ & $\eta_b(2S)$ & $\chi_{b0}(2P)$ & $\eta_b(3S)$\\
\midrule
charge&  0.165(7) & 0.231(7) & 0.3468(8) & 0.109(1) & 0.175(3) & 0.2235(4) & 0.267(3) & 0.306(2) \\
mass  &   0.150(6) & 0.209(6) & 0.332(2)  & 0.107(4) & 0.171(1) & 0.221(2)    & 0.263(1) & 0.302(2) \\ 
\bottomrule
\end{tabular}
\end{table}

\section{Summary and Outlook}

The light-front Hamiltonian formalism provides a natural framework to 
understand strongly interacting relativistic bound states. We demonstrate 
this in a model that implements a holographic confinement
and a one-gluon exchange interaction in the heavy quarkonium systems. 
We update the previous results by incorporating the evolution of the strong
interaction coupling constant. 
The resulting spectroscopy is compared favorably with the experiments. 
We present the wave functions, which bridge the nonrelativistic picture
and the quantum field theoretical treatment of heavy quarkonium. They 
offer a unique way to visualize and to understand the structure of the relativistic
bound states. 

Our model can be applied to other mesons, e.g., heavy-light and light mesons, 
which is in progress. Another important improvement is to include the 
self-energy correction \cite{Karmanov04, Karmanov07}, which will allow us
to go beyond the light-front potential model treatment and to build a consistent
effective light-front model for hadrons. 

There are several promising ways under development to go beyond the current 
phenomenological model. The first one is the systematic Fock sector expansion with
a non-perturbative sector dependent renormalization \cite{Karmanov08, Li15}. 
The second one is the full configuration interaction simulation, with or without the coupled
cluster technique, of the light-front QCD Hamiltonian within the basis function approach 
\cite{Vary10}.  The third line of investigation is to use the effective Hamiltonians obtained 
from a similarity renormalization group with a systematic perturbative expansion \cite{Glazek01}. 
Last but not least, progress in the holographic approach to QCD may also provide
new insights. 
The current model can be linked to all these lines of investigation. 
The physics learned from the current model, and the techniques employed in our model, 
may be helpful to explore those more ambitious approaches. 

%\newpage 

%%%%%%%%%%%%%%%%%%%%%%%%
\begin{acknowledgements}
I wish to thank the organizers of the Light Cone 2016 conference. I acknowledge valuable discussions with 
J.P. Vary, P. Maris, K. Tuchin, S.J. Brodsky, S.D. G\l{}azek, 
X. Zhao, G. Chen, M. Gomez-Rocha, and S. Leit\~ao. I also want to thank the hospitality 
of the High Energy Nuclear Theory Group at the Institute of Modern Physics, Chinese Academy of
Sciences, Lanzhou, China, where the work is being completed. 
This work was supported in part by the Department of Energy under
Grant Nos.~DE-FG02-87ER40371 and DESC0008485 (SciDAC-3/NUCLEI).
Computational resources were provided by the National Energy Research
Supercomputer Center (NERSC), which is supported by the Office of
Science of the U.S. Department of Energy under Contract No.~DE-AC02-05CH11231.

\end{acknowledgements}
%%%%%%%%%%%%%%%%%%%%%%%%

\appendix
\section{Derivation of Equations (\ref{eqn:light-front_charge_radii}--\ref{eqn:light-front_gravitational_radii})}\label{app}

 We used the identity:
\begin{equation}
 \nabla_{q_\perp}^2 = \frac{\partial^2}{\partial q_\perp^2} + \frac{d-1}{q_\perp} \frac{\partial}{\partial q_\perp} -
\frac{1}{q_\perp^2}L^2_z 
= 4t \frac{\partial^2}{\partial t^2} + 2d \frac{\partial}{\partial t} - \frac{1}{t}L^2_z
\end{equation}
where $d=2$ is the spatial dimension, and $t \equiv q^2_\perp$. The angular momentum operator $L_z = i{\partial}/{\partial\phi}$.
The two-dimensional Laplacian has a different coefficient from the three-dimensional one, which results the 
factor 3/2.
At $t \to 0$, the first term $4t \partial^2/\partial t^2$ vanishes. The form factor does not have angle dependence, 
so, the third term vanishes as well. 
Then, at $t \to 0$, 
\begin{equation}
\langle r^2 \rangle = -6 \frac{\partial }{\partial t} F(t)\Big|_{t\to 0} =-\frac{3}{2} \nabla^2_{q_\perp} I_{\lambda\lambda'}(\vec q_\perp) \Big|_{q_\perp \to 0},
\end{equation}
where $I_{\lambda\lambda'}(\vec q_\perp)$ is the helicity amplitude. In the LFWF representation \cite{Brodsky00},
\begin{align}
 I^{\mathrm{em}}_{\lambda\lambda'}(\vec q_\perp) =& \sum_{s, s'} \int_0^1\frac{\dd x}{2x (1-x)}\int\frac{\dd^2k_\perp}{(2\pi)^3}
 \psi^{(\lambda')*}_{ss'/h}(\vec k_\perp+(1-x)\vec q_\perp, x) \psi^{(\lambda)}_{ss'/h}(\vec k_\perp, x), \\
 I^{\mathrm{gr}}_{\lambda\lambda'}(\vec q_\perp) =& \sum_{s, s'} \int_0^1\frac{\dd x}{2x (1-x)}\int\frac{\dd^2k_\perp}{(2\pi)^3}
\Big[ x \, \psi^{(\lambda')*}_{ss'/h}(\vec k_\perp+(1-x)\vec q_\perp, x)  \psi^{(\lambda)}_{ss'/h}(\vec k_\perp, x) \nonumber \\
&+  (1-x)\,\psi^{(\lambda')*}_{ss'/h}(\vec k_\perp-x\vec q_\perp, x) \psi^{(\lambda)}_{ss'/h}(\vec k_\perp, x)  \Big].
\end{align}
Equations~(\ref{eqn:light-front_charge_radii}--\ref{eqn:light-front_gravitational_radii}) immediately follow. 

%%%%%%%%%%%%%%%%%%%%%%

\end{document}